\documentclass[fleqn,twoside]{article}
\usepackage{espcrc2}

\usepackage{epsfig}
\usepackage{amsmath}
\usepackage{amssymb}
\usepackage{amsbsy}
\usepackage{amsfonts}
\usepackage{graphics}
\usepackage{latexsym}
\usepackage{graphicx}
\newcommand{\bc}{\begin{center}}
\newcommand{\ec}{\end{center}}
\newcommand{\be}{\begin{equation}}
\newcommand{\ee}{\end{equation}}
\newcommand{\bea}{\begin{eqnarray}}
\newcommand{\eea}{\end{eqnarray}}
\newcommand{\Dlr}{\stackrel{\leftrightarrow}{D}}
\newcommand{\Dl}{\stackrel{\leftarrow}{D}}
\newcommand{\Dr}{\stackrel{\rightarrow}{D}}

\title{
\begin{flushleft}
\vspace*{-1.5cm}
{\normalsize DESY 05-210}\\[-0.3em]
{\normalsize Edinburgh 2005/17}
\vspace*{0.15cm}
\end{flushleft}
Second moment of the pion's distribution amplitude%
\thanks{Talk presented by J.M. Zanotti at Light Cone 2005, Cairns,
  Australia.}
}

\author{
M. G\"ockeler\address{Institut f\"ur Theoretische Physik,
  Universit\"at Regensburg, 93040 Regensburg, Germany},
R. Horsley\address{School of Physics, University of Edinburgh,
  Edinburgh EH9 3JZ, UK},
D. Pleiter\address{John von Neumann-Institut f\"ur Computing NIC /
  DESY, 15738  Zeuthen, Germany},
P.E.L. Rakow\address{Theoretical Physics Division, Department of 
Mathematical Sciences, University of Liverpool,\\ Liverpool L69 3BX,
UK},
A. Sch\"afer$^{\rm a}$,
G. Schierholz$^{\rm c}$\address{Deutsches Elektronen-Synchrotron DESY,
  22603 Hamburg, Germany},
W.~Schroers$^{\rm c}$ and
J.M. Zanotti$^{\rm bc}$ \\
\vspace*{5mm}
\emph{(QCDSF/UKQCD Collaboration)}}

\begin{document}

\begin{abstract}
  We present preliminary results from the QCDSF/UKQCD collaborations
  for the second moment of the pion's distribution amplitude with two
  flavours of dynamical fermions.
  We use nonperturbatively determined renormalisation coefficients
  to convert our results to the $\overline{\rm MS}$ scheme at
  5~GeV$^2$.
  Employing a linear chiral extrapolation from our large pion masses
  $>$~550~MeV, we find $\langle\xi^2\rangle=0.281(28)$, leading to a
  value of $a_2=0.236(82)$ for the second Gegenbauer moment.
\end{abstract}

\maketitle

\section{INTRODUCTION}
\label{sec:intro}
  
The distribution amplitude of the pion, $\phi_\pi (x,\mu^2)$, contains
information on how the pion's longitudinal momentum is divided between
its quark and anti-quark constituents when probed in diffractive dijet
production at E-791 \cite{Aitala:2000hb} and exclusive pion
photoproduction at CLEO \cite{Gronberg:1997fj}.

The pion's distribution amplitude is defined on the light cone as
\begin{eqnarray}
\lefteqn{\langle 0|\bar{d}(0) \gamma_\mu\gamma_5 u(z)|\pi^+ (p)\rangle
  =}   \nonumber\\
&&\qquad i f_\pi p_\mu \int^1_0 dx\ e^{-ixp\cdot z}\phi_\pi (x,\mu)\ ,
\label{eq:lc-da}
\end{eqnarray}
where $z^2=0$ is a vector along the light cone, $x$ is the fraction of
the pion's longitudinal momentum, $p$, carried by the quark
($\bar{x}=1-x$ for the anti-quark), $f_\pi$ is the pion decay constant,
and $\mu$ is the factorisation scale.
On a Euclidean lattice, we are not able to compute matrix elements of
bilocal operators such as $\bar{d}(0) \gamma_\mu\gamma_5 u(z)$ in
Eq.~(\ref{eq:lc-da}), instead we make use of the light-cone operator
product expansion which allows one to calculate Mellin moments of
Eq.~(\ref{eq:lc-da}) via the computation of matrix elements of local
operators.
The $n^{th}$ moment of the pion's distribution amplitude is defined as
\be
\langle\xi^n\rangle \equiv \int\ {\rm d}\xi\, \xi^n\,\phi(\xi,Q^2)\ ,
\ee
where $\xi\equiv x-\bar{x}=2x-1$, and can be extracted from the matrix
elements of twist-2 operators
\be
\langle 0|{\cal O}_{\{\mu_0\ldots\mu_n\}}(0) |\pi(p)\rangle = f_\pi
p_{\mu_0} \ldots p_{\mu_n} \langle\xi^n\rangle + \cdots\ ,
\label{me}
\ee
where
\be
{\cal O}_{\mu_0\ldots\mu_n}(0) = (-i)^n \overline{\psi}
\gamma_{\mu_0}\gamma_5 \Dlr_{\mu_1}\ldots\Dlr_{\mu_n} \psi\ ,
\label{eq:twist2op}
\ee
$\Dlr=\Dr - \Dl$ and $\{\cdots\}$ denotes symmetrisation of indices
and subtraction of traces. 
We implement the standard normalisation by setting
$\langle\xi^0\rangle = 1$. Due to G-parity the first moment,
$\langle\xi^1\rangle$, vanishes for the pion, hence the first
nontrivial moment that we are able to calculate is
$\langle\xi^2\rangle$.

Although the first lattice calculations of $\langle\xi^2\rangle$
appeared almost 20 years ago \cite{Kronfeld:1984zv},
there has been surprisingly little activity in this area in recent
times \cite{Daniel:1990ah,DelDebbio:1999mq,DelDebbio:2002mq} to
complement other theoretical investigations, e.g. 
\cite{Chernyak:1983ej,Braun:1989iv,Braun:2003rp,Ball:1998je,Ball:2003sc,Bakulev:2001pa,Schmedding:1999ap,Bakulev:2004cu,Diehl:2001dg,Bakulev:2002uc,Dalley:2002nj,Ball:2005tb}.
The current state-of-the-art lattice calculation comes from Del Debbio
{\it et al.} \cite{DelDebbio:2002mq} who performed a simulation in
quenched QCD and renormalised their results perturbatively to the
$\overline{\rm MS}$ scheme at $\mu=2.67$~GeV, 
$\langle\xi^2\rangle^{\overline{\rm MS}}(\mu=2.67\,{\rm GeV}) =
0.280(49)^{+0.030}_{-0.013}\ .$

In these proceedings we present preliminary results from the
QCDSF/UKQCD collaborations for $\langle\xi^2\rangle^{\overline{\rm
    MS}}$ in two flavour lattice QCD.
These results complement our preliminary results on the pion form
factor \cite{pionFF}.


\section{OPERATORS}
\label{sec:operators}
  
The $H(4)$ representation on the lattice leads us to consider two
operators which we call ${\cal O}^a$ and ${\cal O}^b$
\cite{Gockeler:2004xb}, e.g.
\bea
\hspace*{-7mm}\vec{p} &=& (p,p,0): \nonumber\\
&\hspace*{-7mm}{\cal O}^{a}_{412}\hspace*{-7mm}& = {\cal O}_{\{412\}} \, ,
\label{eq:Oa412}\\
\hspace*{-7mm}\vec{p} &=& (p,0,0): \nonumber\\
&\hspace*{-7mm}{\cal O}^{b}_{411}\hspace*{-7mm}& = \left({\cal
    O}_{\{411\}} - \frac{{\cal O}_{\{422\}} + {\cal O}_{\{433\}}}{2}
\right)\, ,
\eea
where $p=2\pi/L_s$ and $L_s$ is the spatial extent of our lattice.
From Eq.~(\ref{me}), we see that ${\cal O}^a$ requires two spatial
components of momentum, while ${\cal O}^b$ needs only one.
Consideration of this fact alone would lead one to choose ${\cal O}^b$,
since units of momenta in different directions on the lattice lead to
a poorer signal.
However, lattice operators with two or more covariant derivatives can
mix with operators of the same or lower dimension.
For forward matrix elements, ${\cal O}^b$ suffers from such mixings
while ${\cal O}^a$ does not.
For matrix elements involving a momentum transfer between the two
states, i.e. nonforward matrix elements, both operators ${\cal O}^a$
and ${\cal O}^b$ can mix with operators involving external ordinary
derivatives, i.e. operators of the form $\partial_\mu\partial_\nu
\cdots (\bar{\psi}\cdots\psi)$. For example, ${\cal O}^{a}_{412}$ in
Eq.~(\ref{eq:Oa412}) can mix with the following operator
\cite{Gockeler:2004xb}
\be
{\cal O}_{\{412\}}^{a,\,\partial\partial} = -\frac{1}{4}
\partial_{\{4} \partial_1 \left(\bar{\psi} \gamma_{2\}} \gamma_5
        \psi\right)\ .
\ee
The situation for ${\cal O}^b$ is a lot worse as it can potentially
mix with up to seven different operators \cite{Gockeler:2004xb}.
Hence a complete calculation of $\langle\xi^2\rangle$ would require
knowledge of the mixing coefficients and renormalisation constants for
all of these mixing operators.
It now becomes obvious that ${\cal O}^{a}$ offers the best possibility
to extract a value of $\langle\xi^2\rangle$ from a lattice simulation.

Although the mixing coefficient for ${\cal O}^{a,\,\partial\partial}$
is not yet known, we expect that it is small and hence we anticipate
that the contribution to $\langle\xi^2\rangle$ from ${\cal
  O}^{a,\,\partial\partial}$ will be small.
Hence, for the rest of the work presented here, we will consider only
the contribution from ${\cal O}^a$.
In a forthcoming publication, we will attempt to address all mixing
issues associated with both operators ${\cal O}^a$ and ${\cal O}^b$.

\begin{table}[tb]
\begin{center}
  \begin{tabular}{ccccc}
        $\beta$ & $\kappa_{\rm sea}$ & Volume & $a$~(fm) & 
        $m_{\pi}$~(GeV)     \\ \hline
    5.20 & 0.13420 & $16^3\times 32$ & 0.1226 & 0.9407(19)    \\
    5.20 & 0.13500 & $16^3\times 32$ & 0.1052 & 0.7780(24)    \\
    5.20 & 0.13550 & $16^3\times 32$ & 0.0992 & 0.5782(30)    \\
    5.25 & 0.13460 & $16^3\times 32$ & 0.1056 & 0.9217(20)    \\
    5.25 & 0.13520 & $16^3\times 32$ & 0.0973 & 0.7746(25)    \\
    5.25 & 0.13575 & $24^3\times 48$ & 0.0904 & 0.5552(14)   \\
    5.29 & 0.13400 & $16^3\times 32$ & 0.1039 & 1.0952(18)    \\
    5.29 & 0.13500 & $16^3\times 32$ & 0.0957 & 0.8674(17)   \\
    5.29 & 0.13550 & $24^3\times 48$ & 0.0898 & 0.7180(13)   \\
    5.29 & 0.13590 & $24^3\times 48$ & 0.0856 & 0.5513(20)   \\
    5.40 & 0.13500 & $24^3\times 48$ & 0.0821 & 0.9692(14)   \\
    5.40 & 0.13560 & $24^3\times 48$ & 0.0784 & 0.7826(17)   \\
    5.40 & 0.13610 & $24^3\times 48$ & 0.0745 & 0.5856(22)
\end{tabular}
\vspace*{2mm}
\caption{\emph{Lattice parameters:
 Gauge coupling $\beta$, sea quark hopping parameter $\kappa_{\rm sea}$,
 lattice volume, lattice spacing and
 pion mass.}
  \label{table:parameters}}
\end{center}
\vspace{-5mm}
\end{table}

\section{LATTICE TECHNIQUES}
\label{sec:lattice}
  
We simulate with $N_f=2$ dynamical configurations generated with
Wilson glue and nonperturbatively ${\cal O}(a)$ improved Wilson
fermions.
For four different values $\beta=5.20$, $5.25$, $5.29$, $5.40$ and up
to four different kappa values per beta we have 
generated ${\cal O}(2000-8000)$ trajectories.
Lattice spacings and spatial volumes vary between 0.075-0.123~fm and
(1.5-2.2~fm)$^3$ respectively.
A summary of the parameter space spanned by our dynamical
configurations can be found in Table~\ref{table:parameters}.
We set the scale via the force parameter, with $r_0=0.5$~fm.

Correlation functions are calculated on configurations taken at a
distance of 10 trajectories using 4 different locations of the fermion
source.
We use binning to obtain an effective distance of 20 trajectories.
The size of the bins has little effect on the error, which indicates
residual auto-correlations are small.

We calculate the average of matrix elements computed with three
choices of pion momenta 
$\vec{p}_0 = ( p, p, 0 )$,
$\vec{p}_1 = ( p, 0, p )$,
$\vec{p}_2 = ( 0, p, p )$,
with the indices of the operators (Eq.~(\ref{eq:Oa412})) chosen
accordingly.

\begin{figure}[t]
\bc
\vspace*{-8mm}
\includegraphics[angle=-90,width=1.13\hsize]{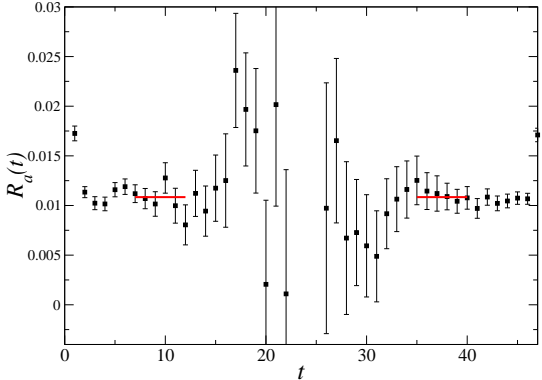}
\vspace*{-10mm}
\caption{\emph{Ratio defined in Eq.~(\ref{eq:ratio-a}), using ${\cal
      O}^a$ from Eq.~(\ref{eq:Oa412}), on a $24^3\times 48$
    lattice with $\beta=5.40$ and $\kappa_{\rm sea}=0.13560$.}}
\label{fig:signal}
\ec
\vspace*{-5mm}
\end{figure}

We define a pion two-point correlation function as
\bea
\lefteqn{C^{\cal O}(t,\vec{p}) = \sum_{\vec{x}} e^{i\vec{p}\cdot\vec{x}}
\left\langle {\cal O}(\vec{x},t) J(\vec{0},0)^\dagger \right\rangle\ ,}
  \nonumber\\
&\rightarrow& \frac{Z}{2E} \langle 0|{\cal O}(0) |\pi(p)\rangle
e^{-Et},\ \ t\gg0\, ,
\eea
where $Z=\langle \pi(p)| J(0)^\dagger |0\rangle$ and we use
$J(x)\equiv\pi(x) = \overline{\psi}(x)\gamma_5\psi(x)$ as the
interpolating operator for the pion. 
The matrix elements in Eq.~(\ref{me}) are then extracted from the
following ratios of two-point functions,
\bea
R^a &=& \frac{C^{{\cal O}^a_{4ij}}(t)}{C^{{\cal O}_4}(t)} 
    = p_i p_j \ \langle \xi^2 \rangle^a \ ,
\label{eq:ratio-a} \\
R^b &=& \frac{C^{{\cal O}^b_{4ii}}(t)}{C^{{\cal O}_4}(t)} 
    = p_i^2 \ \langle \xi^2 \rangle^b \ ,
\eea
where $i$ and $j$ are spatial indices, and ${\cal O}_4$ is the
operator given in Eq.~(\ref{eq:twist2op}) with no derivatives and
$\mu_0=4$.

Figure~\ref{fig:signal} shows a typical example of the ratio in
Eq.~(\ref{eq:ratio-a}) where we clearly observe two plateaus for
$t<L_T/2$ and $t>L_T/2$, where $L_T$ is the time extent of the
lattice.
After extracting $R^a$ from the plateaus, we use
Eq.~(\ref{eq:ratio-a}) to extract $\langle \xi^2 \rangle$.

\begin{figure}[t]
\bc
\vspace*{-11mm}
\includegraphics[angle=-90,width=1.13\hsize]{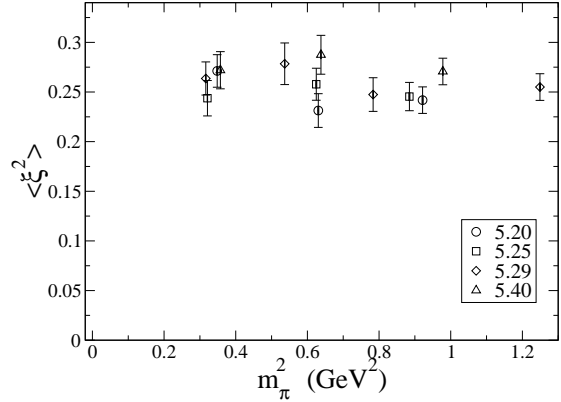}
\vspace*{-7mm}
\caption{\emph{$\langle\xi^2\rangle$ as a function of $m_\pi^2$ for
    ${\cal O}^a$ from Eq.~(\ref{eq:Oa412}).}}
\label{fig:xi2-mpi2}
\ec
\vspace*{-5mm}
\end{figure}

In general, bare lattice operators must be renormalised in some scheme,
${\cal S}$, and at a scale, $M$,
\be
{\cal O}^{\cal S}(M) = Z^{\cal S}_{\cal O}(M) {\cal O}_{\rm bare}\ ,
\ee
so in order to calculate a renormalised value for $\langle \xi^2
\rangle$, we must consider
\be
\langle \xi^2 \rangle^{\cal S}(M) = \frac{Z^{\cal S}_{\cal O}(M)}
{Z_{{\cal O}_4}} \langle\xi^2\rangle_{\rm bare}\ .
\label{eq:renorm-xi2}
\ee
We choose to renormalise to the $\overline{\rm MS}$ scheme at a scale of
$\mu^2=5\,{\rm GeV}^2$. 
Further details of our renormalisation techniques can be found in
\cite{Gockeler:2004wp} and a forthcoming publication.

\begin{figure}[t]
\bc
\vspace*{-4mm}
\includegraphics[angle=-90,width=1.1\hsize]{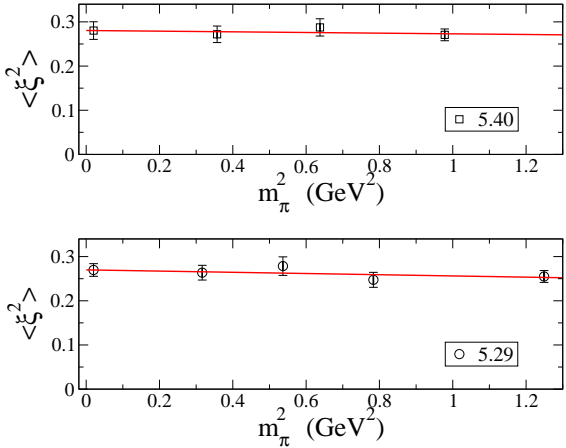}
\vspace*{-6mm}
\caption{\emph{Chiral extrapolation at constant $\beta$ for
    $\beta=5.40$ (top) and  $\beta=5.29$ (bottom) for
    ${\cal O}^a$ from Eq.~(\ref{eq:Oa412}).}}
\label{fig:xi2-chiral}
\ec
\vspace*{-5mm}
\end{figure}

\section{RESULTS FOR $\langle\xi^2\rangle$}
\label{sec:results}
  
For each of our datasets, we extract a value for
$\langle\xi^2\rangle_{\rm bare}$ from Eq.~(\ref{eq:ratio-a}) and
renormalise using Eq.~(\ref{eq:renorm-xi2}).
Figure~\ref{fig:xi2-mpi2} shows these results plotted as a function of 
$m_\pi^2$.
Here we observe that the results are approximately constant as we vary
the pion mass.

\begin{figure}
\bc
\vspace*{-4mm}
\includegraphics[angle=-90,width=1.1\hsize]{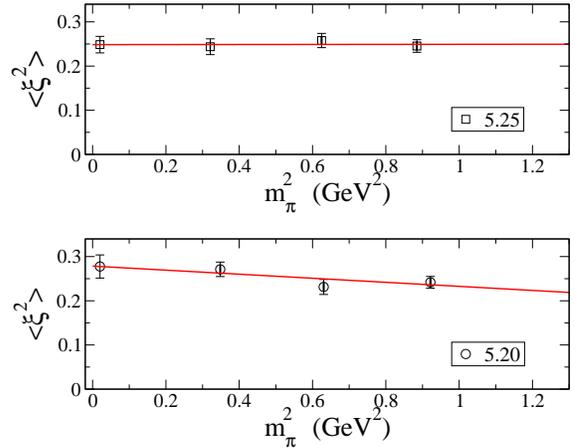}
\vspace*{-6mm}
\caption{\emph{Chiral extrapolation at constant $\beta$ for
    $\beta=5.25$ (top) and  $\beta=5.20$ (bottom) for
    ${\cal O}^a$ from Eq.~(\ref{eq:Oa412}).}}
\label{fig:xi2-chiral2}
\ec
\vspace*{-5mm}
\end{figure}

In order to obtain a result in the continuum and chiral limits, we
first extrapolate our results at constant $\beta$ to the chiral
limit. 
In Fig.~\ref{fig:xi2-chiral}  we display the
chiral extrapolations for $\beta=5.40$ (top) and 5.29 (bottom), while
Fig.~\ref{fig:xi2-chiral2} contains the corresponding extrapolations for
$\beta=5.25$ (top) and 5.20 (bottom).
These results exhibit only a mild dependence on the quark mass and
their values in the chiral limit agree within errors.

\begin{figure}[t]
\bc
\vspace*{-11mm}
\includegraphics[angle=-90,width=1.1\hsize]{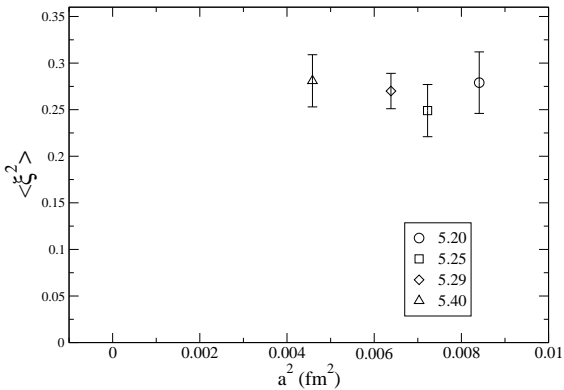}
\vspace*{-6mm}
\caption{\emph{Results for each value of $\beta$ in the chiral limit
    as a function of $a^2$ for
    ${\cal O}^a$ from Eq.~(\ref{eq:Oa412}).}}
\label{fig:xi2-cont}
\ec
\vspace*{-5mm}
\end{figure}

Now that we have calculated results in the chiral limit for each
choice of $\beta$, we are in a position to examine the behaviour of
our results as a function of the lattice spacing.
In Fig.~\ref{fig:xi2-cont} we use the values of $r_0$ calculated in
the chiral limit for each $\beta$ (see Table~3 of
Ref.~\cite{Gockeler:2005rv}) to study the dependence of our results on
the lattice spacing.
Here we observe that even though our operators are not ${\cal
  O}(a)$-improved, we find a negligible dependence on the lattice
spacing, at least when compared to the statistical errors.
Hence we take the result at our smallest lattice spacing (largest
$\beta$) as our result in the continuum limit.

We find in the continuum limit at the physical pion mass, 
the second moment of the pion's distribution amplitude to be
\be
\langle\xi^2\rangle^{\overline{\rm MS}}(\mu^2=5\,{\rm GeV}^2) =
0.281(28)\ ,
\label{eq:xiresult}
\ee
which is very close to the value 
$\langle\xi^2\rangle^{\overline{\rm MS}}(\mu=2.67\,{\rm GeV}) =
0.280(49)^{+0.030}_{-0.013}$ found in Ref.~\cite{DelDebbio:2002mq},
and larger than the asymptotic value, $\mu^2\rightarrow\infty$
\be
\phi_{\rm as}(\xi) = \frac{3}{2}(1-\xi^2)\ \Rightarrow\ 
\langle\xi^2\rangle = 0.2\ .
\ee

\section{GEGENBAUER MOMENT $a_2$}
\label{sec:gegenbauer}

The distribution amplitude, $\phi(x,\mu^2)$, can be expanded in a
series of even Gegenbauer polynomials, $C_{2n}^{\frac{3}{2}} (2x-1)$
\cite{Braun:1989iv,Braun:2003rp},
\be
\phi(x,\mu^2) = 6x(1-x) \sum_{n=0}^\infty a_{2n}(\mu^2) 
C_{2n}^{\frac{3}{2}} (2x-1), 
\label{eq:gegen1}
\ee
where $a_{2n}(\mu^2)$ are the multiplicatively renormalisable Gegenbauer
moments.
Since the higher moments, $n>4$, are expected to be small, it is
common to truncate Eq.~(\ref{eq:gegen1})
\bea
\phi (x,\mu^2) &=& 6x(1-x) \left\{
    1+a_2(\mu^2)C_2^{\frac{3}{2}} (2x-1)\right.  \nonumber \\
&+& \left. a_4(\mu^2)C_4^{\frac{3}{2}} (2x-1)\right\}\, .
\label{eq:gegen2}
\eea
In the asymptotic limit, $\mu^2 \rightarrow \infty$, all
$a_{2n}(\mu^2)=0$, for $n>0$.

Analyses of CLEO  data for $F_{\pi\gamma\gamma^*}$ constrain the pion
distribution amplitude by calculating a relationship between the first
two Gegenbauer moments (see eg., \cite{Diehl:2001dg,Bakulev:2002uc}),
together with upper and lower bounds on their respective values.

Taking the second $\xi=2x-1$ moment of the r.h.s. of
Eq.~(\ref{eq:gegen2}) gives
\be
\langle \xi^2 \rangle = \int^1_0 d\xi\ \xi^2 \phi(\xi,\,\mu^2) =
\frac{1}{35}\left( 7+12a_2(\mu^2) \right)\, ,
\ee
which allows us to extract $a_2$, but not $a_4$.
In order to place a constraint on $a_4$, we would need to calculate
the fourth moment of the pion's distribution amplitude which requires
using an operator involving four covariant derivatives.

Using our result in Eq.~(\ref{eq:xiresult}) we calculate 
\be
a_2(\mu^2=5\,{\rm GeV}^2) = 0.236(82)\ ,
\ee
which is larger than the values $a_2(1\,{\rm GeV}) = 0.07(1)$
\cite{Dalley:2002nj} and $a_2(1\,{\rm GeV}) = 0.19(19)$
\cite{Ball:2005tb}, but within the usual constraints $0\le a_2(1\,{\rm
  GeV}) \le 0.3$ commonly quoted in the literature, even when we
consider that our value will increase slightly when we run our result
to a smaller scale of $\mu=1$~GeV.

\vspace*{2mm}
\section{CONCLUSIONS}
\label{sec:conclusions}
\vspace*{2mm}
  
We have presented a preliminary result for the second moment of the
pion's distribution amplitude calculated on lattices generated by the
QCDSF/UKQCD collaboration with two flavours of dynamical fermions.
We use nonperturbatively determined renormalisation coefficients to
convert our result to the $\overline{\rm MS}$ scheme at 5~GeV$^2$.
We find $\langle\xi^2\rangle=0.281(28)$, which is very close to 
an earlier lattice result and larger than the asymptotic value.

Using a fourth order Gegenbauer polynomial expansion, we calculate a
value for the second Gegenbauer moment, $a_2(\mu^2=5\,{\rm GeV}^2) =
0.236(82)$.

Although we have only employed a linear chiral extrapolation and our
operators are not ${\cal O}(a)$-improved, the chiral and continuum
extrapolations do not seem to be a major source of systematic error
when compared to the statistical errors.
These issues will be addressed in more detail in a forthcoming coming
publication, where we also intend to investigate finite size and
(partially) quenching effects as well as renormalisation group running
of the relevant matrix elements.


\vspace*{2mm}
\section*{ACKNOWLEDGEMENTS}
\vspace*{2mm}

The numerical calculations have been done on the Hitachi SR8000 at LRZ
(Munich), on the Cray T3E at EPCC (Edinburgh) \cite{UKQCD} and on the
APE{\it 1000} at DESY (Zeuthen). 
This work was supported in part by the DFG (Forschergruppe
Gitter-Hadronen-Ph\"anomenologie) and by the EU Integrated
Infrastructure Initiative Hadron Physics (I3HP) under contract
RII3-CT-2004-506078.


\end{document}